\documentclass[useAMS,usenatbib]{mn2e}
\usepackage{graphicx,psfrag}
\usepackage{color}
\begin{document}


\title[Shock formation in stellar perturbations and tidal shock waves
  in binaries]{Shock formation in stellar perturbations and tidal shock waves
  in binaries}

\author[Carsten Gundlach and Jeremiah W. Murphy]
{Carsten Gundlach$^{1}$ \thanks{Email: cjg@soton.ac.uk}
and Jeremiah W. Murphy$^{2}$ \\ 
$^{1}$School of Mathematics, University of
  Southampton, Southampton, SO17 1BJ, UK \\ 
$^{2}$Astronomy Department,
  University of Washington, Box 351580, Seattle, WA 98195-1580 \\ 
(NSF
  Astronomy and Astrophysics Postdoctoral Fellow)}

\date{24 March 2011, revised 23 May 2011}

\maketitle


\begin{abstract}

We investigate whether tidal forcing can result in sound waves
steepening into shocks at the surface of a star.  To model the sound
waves and shocks, we consider adiabatic non-spherical perturbations of
a Newtonian perfect fluid star.  Because tidal forcing of sounds waves
is naturally treated with linear theory, but the formation of shocks
is necessarily nonlinear, we consider the perturbations in two
regimes.  In most of the interior, where tidal forcing dominates, we
treat the perturbations as linear, while in a thin layer near the
surface we treat them in full nonlinearity but in the approximation of
plane symmetry, fixed gravitational field and a barotropic equation of
state. Using a hodograph transformation, this nonlinear regime is also
described by a linear equation. We show that the two
regimes can be matched to give rise to a single mode equation which is
linear but models nonlinearity in the outer layers. This can then be
used to obtain an estimate for the critical mode amplitude at which a
shock forms near the surface. As an application, we consider the tidal
waves raised by the companion in an irrotational binary system in
circular orbit. We find that shocks form at the same orbital
separation where Roche lobe overflow occurs, and so shock formation is
unlikely to occur.

\end{abstract}



\section{Introduction}

As far as we are aware it is unknown if the tidal forces in a binary
inspiral can create shock waves before the binary objects touch, begin
mass transfer or plunge. In order to investigate this, we have
developed a quantitative criterion for the critical amplitude at which
stellar perturbations form shocks that may be interesting in its own
right, or in other applications.

This work was originally motivated by the observation of unsmooth
fluid behaviour in the numerical simulation of an irrotational, equal
mass neutron star (NS) binary merger (\cite{RezzollaBNS1,
  RezzollaBNS2}). The simulations show surface waves breaking when the
initial data are evolved with the cold equation of state (EOS)
$P=K\rho^2$ , and a strong wind when they are evolved with the
equivalent hot EOS $P=\rho e$ (with initially constant entropy). As
both simulations should be identical until genuine shocks form, it
seems likely that both the wind and the surface waves are artefacts of
the interaction with the artificial atmosphere. On the other hand,
these artefacts may also hide genuine shocks.

Mass shedding in small amplitude nonlinear perturbations has been
demonstrated numerically for neutron stars rotating near the
mass-shedding angular velocity
(\cite{StergioulasApostolatosFont,DimmelmeyerStergioulasFont}), but the
minimum perturbation amplitude for this to occur has not been
quantified.

We have therefore tried to obtain a quantitative criterion for shock
formation using a combination of stellar perturbation theory and
nonlinear planar fluid dynamics. We consider a shock formation
scenario where essentially radial sound waves steepen as they approach
the surface because the density and sound speed approach zero at the
surface. (Note that while the shape of the tidal bulges rotates around
the star, individual fluid elements mainly move up and down.) 

If such waves are generated by tidal forces from the companion, their
amplitude and shape is determined in the bulk of the star, where
almost all the mass is. In this regime, linear perturbation theory can
be used to obtain the response of the star to the tidal force,
treating its proper oscillation modes as forced harmonic oscillators.
For simplicity, we assume that the background star is irrotational and
spherical.

On the other hand, near the surface the fluid geometry can be
approximated as plane-parallel, and entropy or composition gradients
become irrelevant compared to the density gradient. In this regime we
use a hodograph transform to cast the nonlinear dynamics into a single
{\em linear} second-order PDE. A shock forms if and only if the
hodograph transform becomes singular: a criterion for this can be
examined within the model itself (\cite{GundlachPlease}). This criterion
was tested in the numerical evolution of nonlinear spherical
perturbations of an $n=1$ polytropic star and found to be accurate
within 10\% (\cite{GablerSperhakeAndersson}). Once formed, these shocks
quickly take a universal, self-similar form (\cite{GundlachLeveque}).

The two regimes are linked by noting that under certain
approximations the fluid variables and their equations of motion in
the two regimes coincide near the surface.

In Sec.~\ref{section:perturbations}, we derive the combination of
linear non-spherical and planar non-linear fluid motion.  We summarise
the (well-known) linear perturbation equations for adiabatic
non-spherical stellar oscillations in a suitable notation, and their
limit near the surface if the density vanishes there. We present the
hodograph transform and the shock formation criterion. We cast the
hodograph equation in a form where it can be identified with the
linear perturbation equations near the surface in a high-frequency
Cowling approximation. We then evaluate the shock formation criterion
on solutions of the linear perturbation equations as if they were
solutions of the hodograph equation. 

In Sec.~\ref{section:tidal}, we apply this general formalism to waves
raised in a star by the tidal force of its binary companion. We can
then use standard methods to calculate the reaction to this force by
expanding the perturbations in proper oscillation modes. We obtain the
orbital separation $d_{\rm crit}$ at which shocks first form as
function of the modes of the star and the mass ratio $q$.  

In Sec.~\ref{section:polytropes}, we carry out the necessary numerical
mode calculations for stars with polytropic equations of state.
Sec.~\ref{section:conclusions} reviews our main approximations and
states our astrophyical conclusions.

A similar calculation to our Sec.~\ref{section:tidal} has been
carried out for $g$-modes in NSs by \cite{Lai}. Their
frequency is lower than the orbital frequency at merger, and so the
orbital frequency moves through resonance as the orbit shrinks, and
the full time-dependent driven oscillator problem must be
considered. It was assumed that no shock forms, and dissipative
heating was estimated instead. It turns out that the duration of the
resonance is too short to give rise to significant heating. By
contrast, we focus on $p$-modes, which have higher frequencies and are
never in resonance, and estimate their amplitude adiabatically in the
approximation of a stationary circular orbit.  


\section{Nonlinear extension of linear perturbation modes}
\label{section:perturbations}


\subsection{Linear adiabatic perturbation equations} 

We consider linear adiabatic perturbations of a spherically symmetric
static perfect fluid star in Newtonian physics, in the frequency
domain. The background is assumed to be non-rotating and in
hydrostatic equilibrium.  $d/dr$ is denoted by a prime.  The background
quantities are the density $\rho(r)$, pressure $P(r)$, gravitational
potential $\phi(r)$, gravitational acceleration $\phi'(r)\ge0$,
sound speed $c(r)$ defined by $c^2\equiv(\partial P/\partial \rho)_s$,
entropy per rest mass $s(r)$, and Brunt-V\"ais\"al\"a frequency  $N(r)$
defined by
\begin{equation}
-{N^2\over {\phi'}}\equiv{\rho'\over\rho}-{P'\over c^2\rho}.
\end{equation}
The equations of hydrostatic equilibrium for the spherical background
star are
\begin{eqnarray}
P'+{\phi'}\rho=0, \\
\phi''+{2\over r}\phi'=4\pi G\rho.
\end{eqnarray}

The displacement vector of the (polar) nonspherical perturbation is
expanded in spherical harmonics as 
\begin{equation}
\label{xirhdef}
{\vec \xi}({\vec r},t)\equiv e^{-i\omega t}
\ \left[\xi_r(r)Y_{lm}(\theta,\varphi){\vec e}_r +\xi_h(r)r{\vec
    \nabla}_\perp Y_{lm}(\theta,\varphi)\right],
\end{equation}
where 
\begin{equation}
{\vec \nabla}_\perp={1\over r}{\vec e_\theta}{\partial\over\partial\theta}
+{1\over r\sin\theta}{\vec e_\varphi}{\partial\over\partial\varphi}.
\end{equation}
Because the equations are linear, it is customary to make $\vec\xi$
complex as above for ease of calcuation. The physical displacement is
its real part
\begin{equation}
\vec\xi_{\rm real}={\rm Re}\, \vec\xi.
\end{equation}
We neglect axial displacements, which in a
non-rotating star have no restoring force, and are zero modes. The
(real) fluid velocity is simply the time derivative of the displacement, or
\begin{equation}
\label{vecvdef}
{\vec v}_{\rm real}({\vec r},t)={\rm Re}[-i\omega{\vec \xi}({\vec r},t)].
\end{equation}
The Lagrangian perturbation $\delta f$ of any background quantity
$f(r)$ is related to the Eulerian perturbation $f_1$ by
\begin{equation}
\delta f\equiv f_1+{\vec \xi}\cdot{\vec\nabla} f=f_1+\xi_r f'.
\end{equation}
All scalar perturbations are also expanded in spherical harmonics, for
example the Eulerian density perturbation
\begin{equation}
\rho_1({\vec r},t)=e^{i\omega t} \rho_1(r) Y_{lm}(\theta,\varphi).
\end{equation}
The assumption of adiabatic perturbations
is that 
\begin{equation}
\delta P=c^2\delta\rho. 
\end{equation}
For given spherical harmonic index $l$, we define the shorthand
\begin{equation}
F(r)\equiv 1-{l(l+1)c^2\over \omega^2r^2}.
\end{equation}

The radial and (polar) horizontal parts of the Euler equation and the mass
conservation equation can be combined to give a single second-order
ODE for $\xi_r$. With a later approximation in mind, we write this as
\begin{equation}
\label{xireqn}
\xi_r''+A\xi_r'+B\xi_r=S,
\end{equation}
where
\begin{eqnarray}
A &=& 2{c'\over c}-{F'\over F}-{N^2\over g}+{2\over r}-{{\phi'}\over
  c^2}, \\
B &=& \left(2{c'\over c}-{F'\over F}-{N^2\over {\phi'}}\right)
\left({2\over r}-{{\phi'}\over c^2}\right) \nonumber \\
&& -{2\over r^2}-\left({{\phi'}\over
  c^2}\right)'+{F\over c^2}(\omega^2-N^2), \\
\label{Sdef}
S &=& {\phi_1'\over c^2}
-{l(l+1)\over\omega^2r^2}
\left({N^2\over {\phi'}}-2{c'\over c}+{F'\over F}+{2\over
  r}\right)\phi_1.
\end{eqnarray}
This is complemented by the perturbed Poisson equation
\begin{equation}
\label{Poisson1}
\phi_1''+{2\over r}\phi_1'-{l(l+1)\over r^2}\phi_1=4\pi G \rho_1.
\end{equation}
Its source term is the Eulerian density perturbation
\begin{equation}
\rho_1=-{\rho\over F}\left[\xi_r'+\left({2\over r}-{{\phi'}\over
    c^2}-F{N^2\over {\phi'}}\right)\xi_r
-{l(l+1)\over\omega^2r^2}\phi_1\right].
\end{equation}
Finally, the complete perturbation can be reconstructed using
\begin{equation}
\xi_h(r)={1\over F\omega^2r}\left\{-c^2\left[\xi_r'+\left({2\over r}-{{\phi'}\over
    c^2}\right)\xi_r\right]+\phi_1\right\}.
\end{equation}
Our second-order equations for $\xi_r$ and $\phi_1$ can be derived
from the first-order systems given in, for example, \cite{Unno} and
\cite{CDM}. 


\subsection{Expansion near the surface and boundary conditions}
\label{surface}

The boundary conditions for $\phi_1$ are $\phi_1\sim r^l$ at $r=0$ and
$\phi_1'+(l+1)\phi_1/r=0$ at $r=R$. The boundary condition for
$\xi_r$ at $r=0$ is $\xi_r\sim r^{l-1}$ (\cite{Unno}). To find the
boundary condition for $\xi_r$ at $r=R$, we need to expand the
equations to leading order in $x\equiv r-R\le 0$. In the following, $O(x)$
will be shorthand for $O(|x|/R)$. 

We assume that near the surface the EOS is approximated
by the Gamma-law EOS
\begin{equation}
\label{Gammalaw}
P(\rho,e)={\rho e \over n},
\end{equation}
where $P$ is the pressure, $\rho$ is the mass density, $e$ the
internal energy per rest mass, and $n$ a constant. From the first law
of thermodynamics, this is equivalent to 
\begin{equation}
P(\rho,s)=K(s)\rho^{1 + {1\over n}},
\end{equation}
where $s$ is the entropy per rest mass. The form of $K(s)$ does not
matter for our purposes (additional input is required to fix it),
and more generally $K$ can be considered a function of both entropy
and composition to take into account stratification effects in NSs.

For a stratified background star where $K(s)$ is a given
function $K(r)$, we find
\begin{equation}
-{N^2\over {\phi'}}=-{n\over n+1}{K'\over K}.
\end{equation}
In the literature (for example \cite{ShapiroTeukolsky}), a power-law
stratification is often considered which is parameterised by
specifying the equilibrium pressure as
\begin{equation}
P(r)=K_0\rho(r)^{1+{1\over n_0}} \quad \Leftrightarrow \quad
K(r)=K_0\rho^{{1\over n_0}-{1\over n}},
\end{equation}
with $n_0>n$ required for stability. (Note that $n$ characterises the
EOS, $n/n_0$ characterises the stratification, and
$n_0$ appears in the Lane-Emden equation). With $\rho\sim x^{n_0}$ as
$x\to 0$, we have
\begin{equation}
K(r)\sim x^{1-{n_0\over n}},
\end{equation}
and so $K$, $K'$ and $N$ all diverge as $x\to 0$. Clearly this
assumption is unphysical on sufficiently small scales, and so near the
surface the stratification must be adjusted away from a pure power
law.

To make an alternative quantitative assumption, we merely assume that
$N^2$ is finite at the surface, or
\begin{equation}
N^2=O(1).
\end{equation}
In this approximation, the condition of hydrostatic
equilibrium gives
\begin{equation}
\label{c2x}
c^2=-{gx\over n}+O(x^2),
\end{equation}
where
\begin{equation}
g\equiv \phi'(R)
\end{equation}
is the gravitational acceleration at the surface.
It is also clear that 
\begin{equation}
\phi_1=O(1), \quad \phi_1'=O(1).
\end{equation}
We shall also need
\begin{equation}
\phi''=-{2\over r}g+O(x),
\end{equation}
which follows directly from the background Poisson equation assuming
$n\ge 1$. In this limit we have
\begin{eqnarray}
\label{Adef}
A&=&{n+1\over x}+O(1), \\
\label{Bdef}
B&=&-{n\over gx}\omega^2\left[
1-{2(n+1)\over n\sigma^2}+{l(l+1)\over n \sigma^4}\right]
+O(1) \nonumber \\
&\equiv& -{n\over gx}\tilde\omega^2 +O(1), \\
S&=&-{n\over gx}\left[\phi_1'(R)-{l(l+1)\over n\sigma^2}{\phi_1(R)\over
    R}\right]
+O(1) \nonumber \\ &\equiv& -{n\over gx}\tilde S +O(1),
\label{Stilde}
\end{eqnarray}
where 
\begin{equation}
\sigma^2\equiv \omega^2{R\over g} =\omega^2{R^3\over GM},
\end{equation}
is a dimensionless mode frequency. Note that $\tilde\omega$ and
$\tilde S$ are defined as constants. 

Keeping only the leading $O(x^{-1})$ in $A$, $B$ and $S$,
(\ref{xireqn}) near the surface becomes
\begin{equation}
\label{nearsurfaceODE}
{d^2\xi_r\over dx^2}+{n+1\over x}{d\xi_r\over dx}-{n\over
  gx}\tilde\omega^2\xi_r=-{n\over gx}\tilde S.
\end{equation}
The solution of (\ref{nearsurfaceODE}) that is regular at $x=0$ is
\begin{equation}
\label{nearsurfacesoln}
\xi_r(x)=C f_n\left(\sqrt{-{4n\tilde\omega^2x\over g}}\right)
+{\tilde S\over \tilde\omega^2},
\end{equation}
where we have defined the function
\begin{equation}
f_n(z)=2^n \Gamma(n+1)\ z^{-n}J_n(z).
\end{equation}
(This is a regular even function of $z$, normalised to obey
$f_n(0)=1$.) In the full set of equations $\tilde S$ is of course not
given a priori but is itself proportional to $\xi_r$ via (\ref{Sdef})
and (\ref{Poisson1}). We also need to fix an overall factor in the
mode, and we choose to make the mode $\xi$ dimensionless and set
\begin{equation}
\xi_r(R)=1.
\end{equation}
Then $C$ has a definite value (for any given mode and polytropic index
$n$), which is determined by solving the full equations (\ref{xireqn})
and (\ref{Poisson1}). (In the Cowling approximation, where
$\phi_1\equiv 0$, we would have $C=1$.)

The regular solution
(\ref{nearsurfacesoln}) obeys the boundary condition
\begin{equation}
\xi_r'-{n \tilde\omega^2\over (n+1)g}\left(\xi_r-{\tilde S
\over \tilde\omega^2}\right)=0.
\end{equation}
This boundary condition is equivalent to Eq.~(17.69) of
\cite{Cox}, {and the boundary conditions derived in \cite{CDM}}
but is {\em not} equivalent to $\delta P/P=0$. This latter boundary
condition is derived in \cite{Unno} under the assumption
of finite sound speed at the surface, see their Eq.~(18.31), and is
therefore not applicable here.

We introduce the dimensionless radius and mode frequency
\begin{equation}
s\equiv {r\over R}, \qquad 
\sigma^2_\alpha\equiv{R^3\over GM} \omega_\alpha^2.
\end{equation}
(Later, when we consider binaries, $R$ and $M$ will refer to $R_1$ and
$M_1$.)  We can then write the approximation near the surface as
\begin{equation}
\label{nearsurfacesolnbis}
\xi_r(s)=C f_n\left(\sqrt{4n\tilde\sigma^2(s-1)}\right)
+{\tilde S\over \tilde\omega^2}.
\end{equation}


\subsection{Nonlinear isentropic perturbations in the constant
  gravitational field, plane-parallel approximation}
 
In \cite{GundlachPlease}, we considered {\em nonlinear} smooth
adiabatic motions in the approximations of planar geometry, the
barotropic EOS 
\begin{equation}
\label{polytropic}
P=K\rho^{1 + {1\over n}},
\end{equation}
with $K$ constant, and constant gravitational acceleration $g$, and
derived the linear partial differential equation
\begin{equation}
\label{vhodo}
v_{\lambda\lambda}=v_{\mu\mu}+{2n+1\over\mu}v_\mu,
\end{equation}
in the independent variables 
\begin{equation}
\label{munudef}
\mu\equiv 2nc, \qquad \lambda\equiv v+gt.
\end{equation}
Here suffices denote partial derivatives. The surface is now at
$\mu=0$, and the interior of the star at $\mu>0$. The boundary
condition $v_\mu=0$ at $\mu=0$ selects the regular solution.

The criterion for the nonlinear fluid equations to form a shock is
that the transformation from $(x,t)$ to $(\mu,\lambda)$ becomes singular. We
showed that this is equivalent to
\begin{equation}
\label{Deltaval}
\left(1-v_\lambda\right)^2-v_\mu^2<0.
\end{equation}
(Obviously, $v$ must be real in this formula.)
If and only if this condition is obeyed, a shock has formed, and the
solution of (\ref{vhodo}) no longer has physical significance.


\subsection{Matching the two approximations}

We now have two sets of approximation: in the ``perturbation
approximation'', everything is linearised around a spherical
equilibrium solution. In the ``hodograph approximation'', the vertical
fluid motion is treated in full nonlinearity, but we neglect
horizontal motion, entropy gradients, and angular dependence, and
approximate the gravitational field as fixed and constant in space and
time.

We expect that there is an overlap region just below the surface 
where both sets of approximations hold at the
same time. In that region, we should then find the same equation of
motion. To see this, note that the perturbation equations can be
adapted to plane-parallel motion by formally setting $l=0$ and $1/R=0$
(and hence $F=1$ and $\sigma=0$), and to a constant gravitational
field by setting $\phi'(r)=g$ and $\phi_1=0$. Neglecting entropy
gradients corresponds to setting $N^2=0$ and $c^2=-ngx$ (with $g$
constant). With all these approximations, (\ref{xireqn}) reduces to
\begin{equation}
\label{xirapprox}
{d^2\xi_r\over dx^2}+{n+1\over x}{d\xi_r\over dx}-{n\over
  gx}\omega\xi_r=0. 
\end{equation}

We now work from the other side. Consider a real, $\lambda$-periodic
solution of (\ref{vhodo}) of the form
\begin{equation}
\label{lambdaperiodic}
v(\lambda,\mu)={\rm Re}\, (-i\omega) e^{-i\omega\bar t} \bar \xi(\bar
x), 
\end{equation} 
where we have {\em defined}
\begin{eqnarray}
\label{barxtdef}
\bar x \equiv  -{\mu^2\over 4ng},  \qquad
\bar t \equiv {\lambda \over g}.
\end{eqnarray}
Then $\bar \xi(\bar x)$ obeys
\begin{equation}
\label{barxieqn}
{d^2\bar\xi\over \bar x^2}+{n+1\over \bar x}{d\bar\xi\over d\bar x}
-{n\over g\bar x}\omega^2\bar\xi=0,
\end{equation}
which is of course formally the same equation as (\ref{xirapprox}),
although it represents nonlinear physics.  Consider now a solution of
(\ref{vhodo}) that represents a small perturbation about the
hydrostatic equilibrium solution (\ref{c2x}) with $v=0$, in the sense
that
\begin{equation}
|v|\ll c, \qquad |\delta c|\ll c .
\end{equation}
Then from the definitions (\ref{munudef}) and (\ref{barxtdef}) we can
infer that
\begin{equation}
\label{barsim}
  \bar x\simeq x, \qquad \bar t\simeq t.
\end{equation}
Furthermore, identifying the planar velocity $v$ with the radial
velocity $v_r$, comparing (\ref{lambdaperiodic}) with (\ref{vecvdef}),
and using (\ref{barsim}), we have
\begin{equation}
\bar \xi\simeq \xi_r.
\end{equation}
We have now justified the coincidence of (\ref{xirapprox}) and
(\ref{barxieqn}). 

However, the actual limit of the perturbation equations near the
surface is not (\ref{xirapprox}) but (\ref{nearsurfaceODE}). They
differ in that $\tilde\omega$ is not $\omega$, and by the (constant in
$x$) source term $\tilde S$ which is not present at all in the
hodograph approximation. Tracing the differences back to the Euler and
Poisson equations, we see that the terms in (\ref{Bdef},\ref{Stilde})
proportional to $l(l+1)$ arise from horizontal motion, and the middle
term in (\ref{Bdef}) arises from the spherical (rather than planar)
symmetry of the background. The difference between $\tilde\omega$ and
$\omega$ vanishes in the high-frequency limit $\sigma^2\gg 1$.

Generally, the source term $S$ in (\ref{xireqn}) represents the effect
of the perturbed gravitational potential on the fluid displacement. In
the hodograph approximation, such a term cannot be accounted for
because the mathematics require the gravitational field $g$ to be
constant. However, the part $\tilde S/\tilde\omega^2$ of the
near-surface approximation (\ref{nearsurfacesolnbis}) to the mode
$\xi_r$ is constant in space, and so corresponds to the whole
near-surface region bobbing up and down as one. Clearly, this part of
motion has no effect on shock formation. We will therefore identify
$\bar\xi$ with $\xi_r$ minus its constant-in-$x$ part.

(We note in passing that in the high-frequency approximation $\tilde
S/\tilde\omega^2\simeq\phi_1'(R)/\omega^2$. If the mode oscillates
with its own proper frequency, the corresponding displacement
$\phi'_1(R)/\omega^2\cos\omega t$ is precisely what results from 
the gravitational field $-\phi'_1(R)\cos\omega t$.)

In summary, to piece together the two approximations into a single
``nonlinear mode equation'', we solve the standard linear perturbation
equations for $\xi_r$ and $\phi_1$ on the whole domain $0\le r\le R$,
and then consider only the $Cf_n$ part of $\xi_r$ when we evaluate the
shock formation criterion.


\subsection{Evaluating the shock formation criterion for a single mode
with periodic time dependence}

Consider now a mode $\vec\xi_\alpha$ with proper frequency
$\omega_\alpha$ that is driven with another frequency $\omega_f$,
resulting in some dimensionless amplitude $A_\alpha$. (To consider a
mode oscillating freely, we can just set $\omega_f=\omega_\alpha$ in
what follows.) Hence the actual time-dependent displacement radial
displacement is given by
\begin{equation}
\Xi_r(r,\theta,\varphi,t)=A_\alpha R\,  \cos(\omega_f t) \
Y_{lm}(\theta,\varphi)\ \xi_\alpha(r).
\end{equation}
Near the surface this approximates as 
\begin{eqnarray}
\Xi_r(r,\theta,\varphi,t)&\simeq & A_\alpha R \cos(\omega_f t) \
Y_{lm}(\theta,\varphi) \nonumber \\ 
&&\left[C_\alpha f_n\left(\sqrt{-{4n\tilde\omega^2_\alpha x\over g}}\right)
+{\tilde S\over \tilde\omega^2_\alpha}\right].
\end{eqnarray}
From the identification we have discussed, going into the bobbing
frame we obtain
\begin{equation}
\label{vmunu}
v(\lambda,\mu)=- A_\alpha C_\alpha R\, \omega_f Y_{lm} \sin\left({\omega_f\over
  g}\lambda\right) \ f_n\left({\omega_\alpha\over g} \mu\right).
\end{equation}
Here we have, somewhat arbitrarily, chosen to use $\omega_\alpha$ as
an approximation to $\tilde \omega_\alpha$, and $\phi_1'(R)$ as an an
approximation to $\tilde S$. In this formula, we consider $v$ as a
slowly varying function of the angles.

Substituting (\ref{vmunu}) into the shock formation criterion
(\ref{Deltaval}), and first focussing on the $\lambda$-dependence, we
can write the result as 
\begin{equation}
(1-U\cos\tau)^2-(V\sin\tau)^2>0, 
\end{equation}
where
\begin{eqnarray}
y&\equiv &{\omega_\alpha\over g} \mu, \qquad \tau \equiv {\omega_f\over
  g}\lambda, \\
U&\equiv& -A_\alpha C_\alpha Y_{lm} \sigma_f^2 f(y), \\
V&\equiv& A_\alpha C_\alpha Y_{lm} \sigma_f\sigma_\alpha f'(y).
\end{eqnarray}
We find that this criterion is sharpest periodically in $\tau$ when
$\cos\tau=U/(U^2+V^2)$, and hence the criterion over at
least a full oscillation period is equivalent to
\begin{equation}
U^2+V^2>1. 
\end{equation}
Introducing the new shorthands
\begin{eqnarray}
\kappa_\alpha&\equiv&{\sigma_\alpha^2\over\sigma_f^2} ,\\
\psi(n,\kappa,y)&\equiv&\sqrt{f_n(y)^2+\kappa f'_n(y)^2}, \\
\Psi(n,\kappa)&\equiv&\max_y \psi(n,\kappa,y),
\end{eqnarray}
we can write out the shock formation criterion for a single mode as
\begin{equation}
\label{Aalphacrit}
A_\alpha C_\alpha \sigma_f^2 \Psi(n,\kappa_\alpha) 
\max_\theta|Y_{lm}|>1,
\end{equation}
Analysis of the function $f_n$ shows that for $\kappa<2(n+1)$, $\psi(y)$
has its global maximum at $y=0$, so $\Psi=1$. For $\kappa>2(n+1)$, the
global maximum is attained at the point where $f_n+\kappa f_n''=0$, with 
value $\Psi>1$. In particular, if the mode oscillates at its proper
frequency, $\omega_f=\omega_\alpha$, we have $\kappa=1$ and hence
$\Psi=1$. 


\subsection{Evaluating the shock formation criterion for a single mode
with arbitrary time dependence}

Consider now a mode $\vec\xi_\alpha$ driven with an arbitrary
time-dependent amplitude $\Xi_\alpha(t)$, that is
\begin{equation}
\Xi_r(r,\theta,\varphi,t)=R \,\Xi_\alpha(t)\ 
Y_{lm}(\theta,\varphi)\ \xi_\alpha(r).
\end{equation}
Hence, near the surface and in the bobbing frame
\begin{equation}
v=R \, C_\alpha \, \dot\Xi_\alpha\left({\lambda\over g}\right) \ 
Y_{lm}(\theta,\varphi)\ f_n\left({\omega_\alpha\over g}\mu\right).
\end{equation}
Hence we can write in dimensionless form
\begin{equation}
v_\lambda=p'(z)q(y), \qquad v_\mu=p(z)q'(y),
\end{equation}
where
\begin{eqnarray}
p(z) &\equiv& \omega_\alpha^{-1} 
\dot\Xi_\alpha\left({z\over\omega_\alpha}\right), \\
q(y)&\equiv &C_\alpha\,Y_{lm} \sigma_\alpha^2 f_n(y).
\end{eqnarray}
We then have to minimise $(1-v_\lambda)^2-v_\mu^2$ over both $z$
and $y$, to see if it reaches a negative value. 

Note that the periodic case of the previous subsection is recovered
with $p(z)=\sin\tau$, with $\tau=\sqrt{\kappa}z$. 


\section{Perturbations raised by tidal forces}
\label{section:tidal}


\subsection{Calculation of the tidal acceleration}

Consider a binary system with masses $M_1$ and $M_2$ in an elliptic
orbit. Let the orbital angular velocity and the spins of the stars
with respect to an inertial reference system be ${\vec \Omega}$,
${\vec S}_1$ and ${\vec S}_2$. In the (non-inertial) reference system
that moves and spins with star 1 and with origin in its centre of
mass, the Euler equation for star 1 becomes
\begin{equation}
{\partial {\vec {v}}\over\partial t}+({\vec
  {v}}\cdot{\vec\nabla}){\vec {v}} +{{\vec \nabla}P\over
  \rho}+{\vec\nabla}\phi_1 = \vec a, 
\end{equation}
where
\begin{equation}
 \vec a\equiv  -{\vec\nabla}\phi_2 + \ddot {\vec r}_0 \nonumber
 - 2{\vec S}_1 \times {\vec v} -\dot {\vec S}_1 \times {\vec r}
 - {\vec S_1} \times {\vec S_1} \times {\vec r}.
\end{equation}
where ${\vec v}$, $\rho$ and $P$ are the fluid velocity, density and
pressure in star 1, $\phi_1$ and $\phi_2$ are the gravitational
potentials generated by star 1 and star 2, respectively, ${\vec
  r_0(t)}$ is the location of the centre of mass of the binary, and
$\vec e_0(t)$ the unit vector in its direction. Using angular momentum
conservation in the form $d_1^2{\vec \Omega}={\rm const}$, we find
\begin{equation}
\ddot{\vec r}_0=\ddot d_1 {\vec e}_0 
+ {\vec \Omega} \times {\vec \Omega} \times  {\vec r}_0.
\end{equation}

In the following we assume ${\vec S_1}=0$, both because this is
believed to be correct for NS binaries (\cite{BildstenCutler,
  Kochanek}) and because it simplifies the calculation, as we can use
perturbation theory on a spherical background star 1. The distance
between the centre of mass of star 1 and the centre of mass of the
binary is $|{\vec r}_0(t)|\equiv d_1$ and the distance between the
centres of mass of the two stars is $|{\vec r}_2(t)|\equiv d\equiv
d_1+d_2$. Here $M_1d_1=M_2d_2$ by virtue of ${\vec r_0}$ being the
centre of mass.

We approximate $\phi_2$ as spherically
symmetric, that is
\begin{equation}
\phi_2=-{GM_2\over|{\vec r}-{\vec r}_2(t)|},
\end{equation}
and expand ${\vec \nabla}\phi_2$ up to $O(r^2)$ in
${\vec r}$. Then 
\begin{eqnarray}
{\vec a}&=&\left(\ddot d_1+{GM_2\over d^2}-d_1\Omega^2\right)\vec e_0
\nonumber \\
&&-{GM_2\over d^3}{\vec r}
+3{GM_2\over d^3}{\vec r}_{\|0}
+ O(r^2),
\end{eqnarray}
where $\perp\Omega$ denotes the projection into the plane normal to
${\vec \Omega}$ and $\|0$ denotes the projection into the direction
$\vec e_0$. The $O(r^0)$ term in round brackets vanishes by the
assumption that the origin of $\vec r$ is the centre of mass of star
1. The remainder can be written as
\begin{equation}
\vec a=\vec\nabla\chi+O(r^2)
\end{equation}
with 
\begin{equation}
\chi \equiv  - {GM_2\over 2d^3} r^2
+ {3GM_2\over 2d^3}(x\cos\phi+y\sin\phi)^2,
\end{equation}
where we have chosen Cartesian coordinates so that the orbit is in the
$xy$-plane, and where $\phi(t)\equiv \int \Omega\, dt$ is the orbital phase, $\Omega(t)$ the
instantaneous orbital angular velocity, and $d(t)$ the instantaneous
orbital separation.  We can write the tidal potential in terms of
spherical harmonics as
\begin{eqnarray}
\chi &=& -{GM_2\over 2d^3}r^2 
+{3GM_2\over 4d^3}r^2\sin^2\theta
\left[1+\cos2(\phi-\varphi)\right] 
\nonumber \\
&=& {3GM_2\over 4d^3} r^2 \left[\sin^2\theta \, \cos2(\phi-\varphi)
-\left(\cos^2\theta-{1\over 3}\right) \right]
 \nonumber \\
\label{chifinal}
&=& {3GM_2\over d^3} \sqrt{2\pi\over 15} r^2\, {\rm
  Re}\left( e^{-2i\phi}Y_{22}\right) \nonumber \\
&&- {GM_2\over d^3}
\sqrt{\pi\over 5} r^2\, Y_{20} .
\end{eqnarray}
Hence the tidal force is polar, and to leading order in $\vec r$ is
given by quadrupole terms. To linear order in
perturbation theory, the deformations caused by the $Y_{00}$, $Y_{20}$
and $Y_{22}$ terms decouple. In circular orbits, the 
$Y_{20}$ term is time-independent because $d$ and $\Omega$ are
constant, and therefore does not cause shocks. We can neglect it also
for moderately eccentric orbits. The
$Y_{22}$ term is always time-dependent because $\phi$ is: physically,
the tides rotate around the star, so that individual fluid elements
move up and down.


\subsection{Response of the star}

The response of the star to ${\vec a}$
is governed by \cite{Lai}
\begin{equation}
\label{Fma}
\left(\rho{\partial^2\over\partial t^2}+{\cal L}\right){\vec
  \Xi}({\vec r},t) = \rho\, {\vec a},
\end{equation}
where ${\cal L}$ is a linear differential operator containing only
spatial derivatives. 

It is generally assumed (for example, \cite{Unno}) that a star admits
a complete set of eigenmodes which obey
\begin{equation}
{\cal L}\,{\vec \xi}_\alpha({\vec r})
=\rho\, \omega_\alpha^2\,{\vec \xi}_\alpha({\vec r})
\end{equation}
and 
\begin{equation}
\label{basis}
\langle {\vec \xi}_\alpha,{\vec \xi}_\beta\rangle \propto\delta_{\alpha\beta},
\end{equation}
where the inner product is
\begin{eqnarray}
&& \langle {\vec \xi},{\vec \eta}\rangle  \equiv  {1\over M} \int{\vec
  \xi}\cdot{\vec \eta}^*\ \rho r^2\,dr\,d\Omega \\
&& =  {1\over
  M}\sum_{l,m}\int_0^R\left[ \xi^r_{lm} \eta^{r*}_{lm}+
  l(l+1)\xi^h_{lm} \eta^{r*}_{lm}\right]\rho r^2\,dr.
\end{eqnarray}
In the second line, we have assumed a decomposition into spherical
harmonics, and into the radial and horizontal parts defined by
Eq.~(\ref{xirhdef}), and the standard normalisation
$\int|Y|^2\,d\Omega=1$. For simplicity of notation, and because ${\vec
  a}$ is polar, we neglect the axial parts of all vectors. Note that
$\vec\xi$ is by our convention dimensionless, $\vec \Xi$ has dimension
length, and $\vec a$ has dimension acceleration.

Decomposing (\ref{Fma}) into modes, we have
\begin{equation}
\vec \Xi(\vec r,t)=R\, \sum_\alpha \, \Xi_\alpha(t)\, \vec
\xi_\alpha(\vec r),
\end{equation}
where the dimensionless amplitudes
$\Xi_\alpha(t)$ obey
\begin{equation}
\left({d^2\over dt^2}+\omega_\alpha^2\right)\Xi_\alpha={1\over R}
{\langle {\vec \xi}_\alpha,{\vec
    a}\rangle \over\langle {\vec \xi}_\alpha,{\vec \xi}_\alpha\rangle
}\equiv f_\alpha(t).
\end{equation}
The general solution is 
\begin{equation}
\label{xialphaintegral}
\Xi_\alpha(t)= B_\alpha e^{-i\omega_\alpha
  t}+{1\over \omega_\alpha}\int_{t_0}^t
\sin\omega_\alpha(t-s)\,f_\alpha(s)\,ds.
\end{equation}


\subsection{Circular orbits}

For a circular orbit, only the $l=m=2$ component of the tidal
potential is time-dependent, through $\phi(t)=2\Omega t$, where
$\Omega$ is the constant orbital angular velocity. Neglecting the
time-independent parts, and splitting the 3-dimensional gradient into
a horizontal and radial part, we have
\begin{equation}
{\vec a} = {GM_2\over d^3}\sqrt{6\pi\over
    5}{\rm Re}\, e^{-2i\Omega t}
 \left( 2r Y_{22}
  {\vec e}_r+r^2{\vec \nabla}_\perp Y_{22}\right).
\end{equation}
Hence we have 
\begin{equation}
f_{22}(t)={GM_2\over d^3}\sqrt{6\pi\over 5} e^{-2i\Omega t} \,{\langle
  {\vec \xi}_\alpha,(2s,s)\rangle \over\langle {\vec \xi}_\alpha,{\vec
    \xi}_\alpha\rangle }.
\end{equation}
The solution of (\ref{Fma}) is then
\begin{equation}
\label{Xialphaexpr}
{\vec \Xi}({\vec r},t)=R \sum_\alpha\left( B_\alpha e^{-i\omega_\alpha
  t}+ A_\alpha e^{-2i\Omega t}\right)
{\vec \xi}_\alpha({\vec r}), 
\end{equation}
where the dimensionless amplitude of the particular integral is
\begin{equation}
\label{Aalpha}
A_\alpha={1\over \omega_\alpha^2-4\Omega^2}
{GM_2\over d^3}\sqrt{6\pi\over 5} 
\,{\langle {\vec
    \xi}_\alpha,(2s,s)\rangle \over\langle {\vec \xi}_\alpha,{\vec
    \xi}_\alpha\rangle
}.
\end{equation}
Note that in contrast to \cite{Lai}, we approximate the orbit as
circular, and we do not take into account resonance. We assume that
the tidal force is the dominant source of oscillations, and that the
orbit evolves so slowly that transients can be neglected, and so we
set $B_\alpha=0$.

We introduce the binary mass ratio and the dimensionless orbital
separation
\begin{equation}
q\equiv {M_2\over M_1}, \quad \eta\equiv {d\over R_1}.
\end{equation}
We note
\begin{equation}
\Omega_{\rm circ}^2={G(1+q)M_1\over d_{\rm circ}^3}.
\end{equation}
We can then write 
\begin{equation}
\label{Aalphabis}
A_\alpha = \sqrt{6\pi\over 5}\
 {q\over \sigma^2_\alpha\eta^3-4(1+q)}\ 
{\langle
  {\vec \xi_\alpha},(2s,s)\rangle \over\langle
 {\vec \xi_\alpha},{\vec \xi_\alpha}\rangle }.
\end{equation}
Combining this with (\ref{Aalphacrit}), and using $\max_\theta
|Y_{22}|=\sqrt{15/32\pi}$ and $\omega_f=2\Omega$, we have the
dimensionless shock formation criterion 
\begin{equation}
\label{finalcrit}
3 C_\alpha \Psi(n,\kappa_\alpha) {1+q\over \eta^3}  {q\over \sigma^2_\alpha\eta^3-4(1+q)} \
 {\langle
  {\vec \xi_\alpha},(2s,s)\rangle \over\langle
 {\vec \xi_\alpha},{\vec \xi_\alpha}\rangle }>1,
\end{equation}
where $\alpha$ characterises the $l=m=2$ mode that maximises the
criterion, and where
\begin{equation}
\kappa_\alpha={\sigma_\alpha^2 \eta^3\over 4(1+q)}.
\end{equation}


\subsection{Elliptic orbits}
\label{elliptic}

For elliptic orbits, we have
\begin{eqnarray}
f_{22}(t)&=&{GM_2\over d^3(t)}\sqrt{6\pi\over
    5} e^{-2i\phi(t)} \,{\langle {\vec \xi}_\alpha,(2s,s)\rangle
  \over\langle {\vec \xi}_\alpha,{\vec \xi}_\alpha\rangle}, \\
f_{20}(t)&=&- \left({4\Omega^2(t)\over 3}+ {GM_2\over d^3(t)}\right)
\sqrt{\pi\over 5} \,{\langle {\vec \xi}_\alpha,(2s,s)\rangle
  \over\langle {\vec \xi}_\alpha,{\vec \xi}_\alpha\rangle}, \\
f_{00}(t)&=&-{4\Omega^2(t)\over 3} \sqrt{\pi} \,{\langle {\vec \xi}_\alpha,(2s,s)\rangle
  \over\langle {\vec \xi}_\alpha,{\vec \xi}_\alpha\rangle}.
\end{eqnarray}
For slightly elliptic orbits, the time-dependence of $\Omega$ and $d$
is weak, and $l=m=2$ is the dominant time-dependent
deformation. For highly eccentric orbits, both $\Omega^2$ and $d^{-3}$
are sharply peaked at periastron. This has two consequences:
First, depending on how quickly perturbations set up by tidal forces
are damped, it may be appropriate to treat each periastron passage as
a transient, rather than as part of periodic excitation. Secondly,
$f_{22}$, $f_{20}$ and $f_{00}$ are now all comparably
time-dependent. In fact, 
\begin{equation}
\Omega_{\rm pa}^2=(1+e){G(1+q)M_1\over d_{\rm pa}^3},
\end{equation}
where $e$ is the eccentricity and $d_{\rm pa}$ the periastron orbital
separation. The three overlap integrals are also likely to be
comparable.  Moreover, as the orbital frequency even at periastron is
substantially lower than the mode frequency, we have approximately
$\Xi_\alpha(t)\simeq \omega^{-2} f_{\alpha}(t)$, as for the circular
orbit case. Hence we expect even the highly eccentric case to excite
the $22$, $20$ and $00$ perturbations at similar amplitudes
$\Xi_\alpha(t)$, comparable to $\Xi_{22}$ in a circular orbit at
periastron radius.


\subsection{Roche lobe overflow and resonance}

Our shock formation criterion becomes irrelevant once Roche lobe
overflow occurs (or when two stars of identical mass and size touch).
Defining
\begin{equation}
\lambda\equiv {d\over d_{11}}
\end{equation}
where $d_{11}$ is the distance from the centre of star 1 to the
Lagrange point $L_1$, a simple calculation gives the quintic
\begin{equation}
\lambda^5-2\lambda^4+\lambda^3-(1+3q)\lambda^2+(2+3q)\lambda-(1+q)=0.
\end{equation}
The one real solution gives $L_1$ (the four complex solutions give the
other Lagrange points in the complex plane). Approximating the Roche
lobe as a sphere centered on star 1, Roche lobe overflow occurs at
$\eta=\lambda$.  As $q\gg 1$, $\lambda\simeq (3q)^{1/3}$.

Resonance is obtained at an orbital
separation
\begin{equation}
\label{dres}
\eta_{\rm res}= \sigma^{-2/3}\left(4+4q\right)^{1/3}.
\end{equation} 
Our result (\ref{finalcrit}) has been obtained under the assumption that
$\omega_\alpha>2\Omega$, and this is always the case until shock
formation or Roche lobe overflow. 


\section{Mode results for polytropic stars}
\label{section:polytropes}

We have used a publicly available code (\cite{code, christensen96}) to calculate
mode functions $\xi(s)$ and frequencies $\sigma$ and hence determine
the overlap integrals and critical values of $\eta$ as a function of $q$.

We have carried out the calculation for the isentropic ($n_0=n$)
Gamma-law EOS 
\begin{equation}
\label{PKrho}
P=K\rho^{1 + {1\over n}}
\end{equation}
with $n=1/2$, $1$, $3/2$ and $3$. $n=1$ is often used as an
approximate equations of state for cold NS matter. {\bf $n=1/2$
  represents possible stiff neutron star equations of state.} 
$n=3/2$ and $n=3$ are good approximations to non-relativistic and
relativistic degenerate electron pressure, respectively. While these
equations of state are simplistic, they have the advantage that the
resulting stellar models, and hence our results, depend only on $n$,
not on $M_1$ and $R_1$. The mass, radius and polytropic constant are
related by the scaling relation (\cite{ShapiroTeukolsky})
\begin{equation}
R_1^{n-3}\propto K^n M_1^{n-1}.
\end{equation}
In these simple stellar models there is no stratification and hence
there are no $g$-modes. 

Our numerical results for circular orbits with equations of state
$n=1/2$, $1$, $3/2$ and $3$ are shown in the figures. For the first several
modes, we plot the critical value of orbital separation $\eta$ against
mass ratio. In all cases, shock formation first occurs for the lowest
frequency mode, so that is the curve that matters for shock
formation. On the same plot, we also show the orbital separation
$\eta$ at which Roche lobe overflow starts. For all three equations of
state, for all values of $q$, the critical orbital separation for the
formation of tidal shocks coincides, within our approximations, with
the critical orbital separation for Roche lobe overflow.

\begin{figure}
\includegraphics[width=8cm]{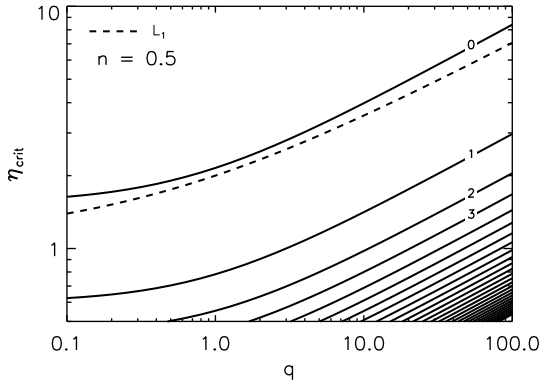}
\caption{ 
\label{figure:eta0.5}
Critical value for shock formation of dimensionless orbital separation
$\eta=d/R_1$ against mass ratio $q=M_2/M_1$. This applies for a
circular orbit, the polytropic equation of state with $n=1/2$
($\Gamma=3$), and $l=m=2$ perturbations. The labels $0,1,2,\dots$
denote the number of radial nodes in the mode. The critical value of
$\eta$ for Roche lobe overflow through the Lagrange point $L_1$ is
shown as a dashed line. At large $q$, $\eta\propto q^{1/3}$ for all curves.}
\end{figure}

\begin{figure}
\includegraphics[width=8cm]{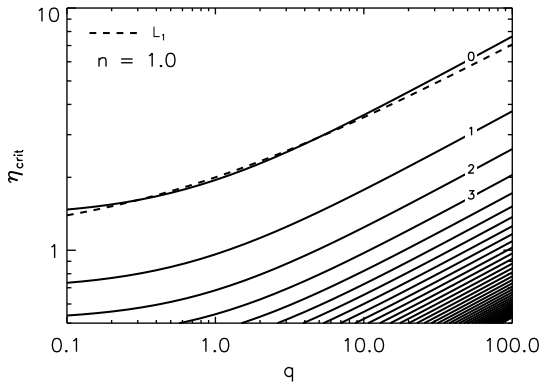}
\caption{ 
\label{figure:eta1.0}
The same plots for the polytropic equation of state with $n=1$
($\Gamma=2$).  }
\end{figure}

\begin{figure}
\includegraphics[width=8cm]{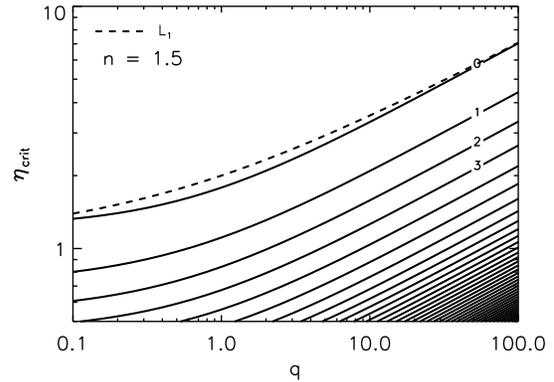}
\caption{ 
\label{figure:eta1.5}
The same plots for the polytropic equation of state with $n=3/2$
($\Gamma=5/3$). }
\end{figure}

\begin{figure}
\includegraphics[width=8cm]{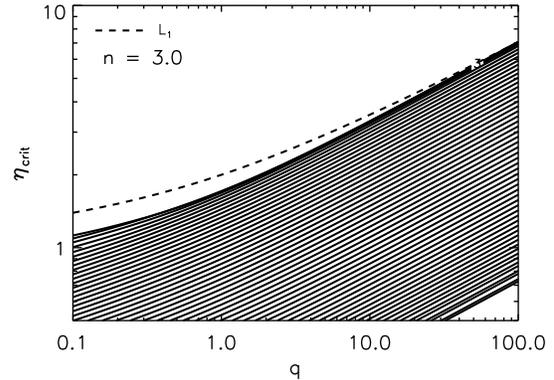}
\caption{ 
\label{figure:eta3.0}
The same plots for the polytropic equation of state with $n=3$
($\Gamma=4/3$). (The bottom right corner appears blank because we have
not calculated modes there. All our figures show the same ranges of
$\eta$ and $q$.)}
\end{figure}

\section{Conclusions}
\label{section:conclusions}

This paper consists of two parts: a quantitative criterion for linear
perturbation modes to form shocks, and an application of this
criterion to modes raised by tidal forces in compact binaries. 

Our shock formation criterion relies on the hodograph transformation
to link linear perturbation modes in the interior to the fully
nonlinear shock formation criterion of \cite {GundlachPlease} near the
surface. This criterion is exact for plane-symmetric motion of a
polytropic fluid in a constant gravitational field. The approximation
of planar symmetry near the surface is natural, and it turns out that
any buoyancy (non-barotropic) effects can also be safely neglected
near the surface as long as the entropy gradient and any composition
gradients are merely bounded. 

Our calculation of the tidal waves in perturbation theory is
straightforward for circular orbits (\cite{Lai}). For stars with a
simple polytropic equation of state $P=K\rho^{1+1/n}$ in irrotational
circular binary orbit, we find that the critical orbital separation
for shock formation essentially coincides with the one for Roche lobe
overflow. {\bf In other words, tidal forces create shocks roughly when
  the binary begins to merge. Within our approximations, the two
  curves agree remarkably closely, so that we cannot say which actually
  occurs first. In any case, the $p$-mode shock formation mechanism we
  have investigated here is not the primary mechanism for binary
  disruption. As discussed above in Sec.~\ref{elliptic}}, we expect the
  same result even for highly elliptic orbits. Although this is a
  negative result, it should be stressed that it was not obvious from
  dimensional analysis: we have also estimated the dimensionless
  factors. 

{\bf Extending our analysis to more realistic stellar
    models would require more extensive modelling, in particular a
    realistic treatment of the surface. (We have shown above in
    Sec.~\ref{surface} that another simple stellar model, assuming one
    polytropic constant for the equation of state and another for the
    stellar structure, gives rise to a divergent Brunt-V\"ais\"al\"a
    frequency, and so is inconsistent with our assumptions of a
    perfect fluid surface). However, the fact that our result holds
    for a polytropic index ranging from $n=1/2$ to $n=3$ suggests that 
    other equations of state would not show shock formation before
    merger either. Intuitively, the weakness of the shock formation
    mechanism is dominated by a factor of (tidal force frequency/mode
    frequency$)^2$ [the factor $\sigma_f^2$ in Eq.~(\ref{Aalphacrit})].}

{\bf In a related result}, \cite{RosswogRamirezHix} give a criterion
for the tidal disruption of a WD in orbit around a much more massive
compact object 2 (black hole or NS) as (in our notation) $\eta_{\rm
  disrupt}\simeq q^{1/3}$ based on numerical simulations. Our results
are consistent with this {\bf for large $q$ in that both Roche lobe
overflow and tidal shock formation occur at the same $\eta$, namely
$\eta_{\rm crit}\simeq 1.44 q^{1/3}$ for $q\gg 1$.}

{\bf Finally, this paper is motivated by the observance of surface
  shocks in numerical simulations of binary neutron star mergers just
  before the stars touch (\cite{RezzollaBNS1, RezzollaBNS2}). Our
  results indicate that these surface shocks are probably not physical.}


\section*{Acknowledgments}

J.W.M. is supported by an NSF Astronomy and Astrophysics Postdoctoral
Fellowship under award AST-0802315.  C.G. would like to thank Randy
Leveque and the University of Washington for hospitality when this
work was begun. We like to thank Eric Agol, Wynn Ho, Tom Maccarone and
John Miller for helpful discussions.


\end{document}